\documentclass[onecolumn,showpacs]{revtex4}
\usepackage{graphicx}
\usepackage{amssymb}
\usepackage{hyperref}

\begin{document}

\title{An archive for quasi-elastic electron-nucleus scattering data}

\date{\today}

\author{Omar Benhar}
\email{benhar@roma1.infn.it}
\affiliation{INFN, Sezione di Roma, I-00185 Roma, Italy \\
Dipartimento di Fisica, Universit\`a ``La Sapienza'',
I-00185 Roma, Italy}

\author{Donal Day}
\email{dbd@virginia.edu}
\affiliation{Dept. of Physics, University of Virginia, Charlottesville,
VA 22903, USA}
\author{Ingo Sick}
\email{ingo.sick@unibas.ch}
\affiliation{Dept. f{\"u}r Physik und Astronomie, Universit{\"a}t Basel,
 CH-4056 Basel, Switzerland}

\begin{abstract}
In connection with a review article on inclusive quasi-elastic electron-nucleus scattering
submitted to Reviews of Modern Physics, we have collected as many cross sections
as possible, and placed them in an web-archive accessible at
\href{http://faculty.virginia.edu/qes-archive}{http://faculty.virginia.edu/qes-archive}. This brief note is intended to alert potential users of these data to this resource.
\end{abstract}
\maketitle
Inclusive quasi-elastic electron-nucleus scattering,  in the course of the last 40 years, has been experimentally investigated at many different facilities. The first data, from light nuclei, were collected in order to extract information about the nucleon elastic form factors. Subsequently investigators have measured quasi-elastic scattering cross sections over a range of momentum and energy transfers. The motivations have been varied and include:   to allow the separation of the nuclear response into its longitudinal and transverse pieces, to extract information about the ground state nuclear momentum distributions, to study possible medium modifications of the nucleon response,  and to study the scaling of the structure functions. Quasi-elastic electron-nucleus scattering data also provides information useful for predicting cross sections for neutrino reactions.

The  cross sections measured over the years have not been, until now, easily accessible. This has been in part due to the large number of data points needed to describe the inclusive continuum. As a consequence, in many cases, the numerical values of the data have not been  published and could only be found in theses, internal reports or the computer-files of some of the original investigators.

\begin{figure}[!h]
   \centering
   \includegraphics[width=\textwidth]{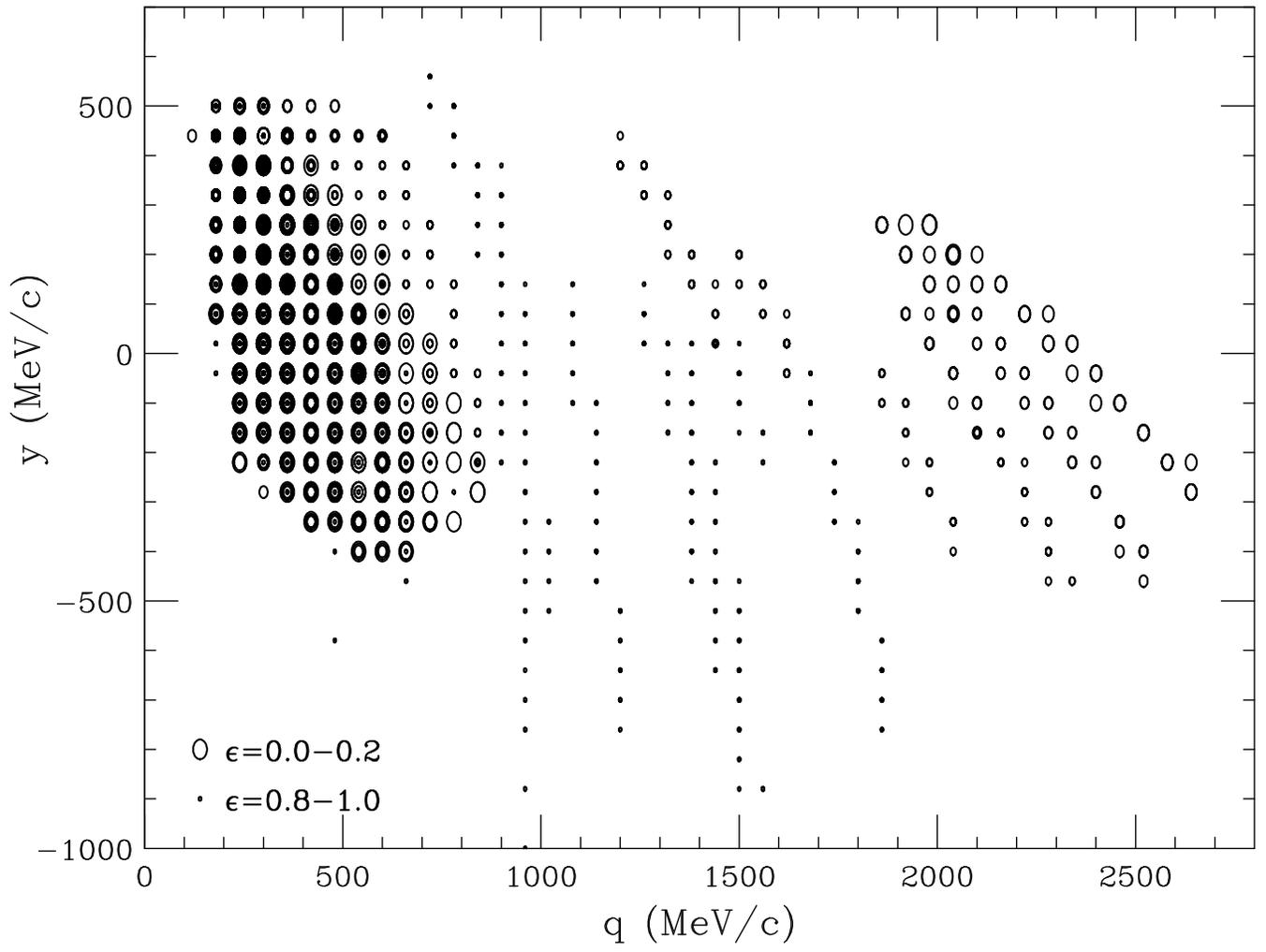} 
   \caption{We show here the kinematic extent of the $^{12}$C data in the archive plotted as function of the scaling variable $y$\ and the momentum transfer $q$.  The symbols indicate the range of $\epsilon$,  the polarization of the virtual photon at each data point, for bin-widths of 0.2.}
   \label{fig:yvsq}
\end{figure}

In connection with a review article\cite{Benhar:2006wy}  on inclusive quasi-elastic electron-nucleus scattering and
submitted to Reviews of Modern Physics, we have assembled all the  quasi-elastic electron-nucleus scattering data that we have succeeded in obtaining, and have made them available to the community via a web-page at \href{http://faculty.virginia.edu/qes-archive}{http://faculty.virginia.edu/qes-archive}. Currently the archive  contains some 18,500 data points.  Figure~\ref{fig:yvsq} presents the kinematic extent of the $^{12}$C data  available.

The reader should be aware that the inclusive cross sections in the quasi-elastic region are not purely so -- they result from  multiple reaction processes, including both quasi-elastic and inelastic scattering from the nucleons. With increasing momentum transfers the contribution from inelastic scattering to the total cross section grows, to the point that it obscures the quasi-elastic peak. Nonetheless we will refer to these data as quasi-elastic.

We are positive that additional data will be located. If these data are passed on to  us we will add them to the archive. 

\end{document}